\documentclass[preprint,prl,showpacs]{revtex4}

\begin{document}

\title{Quantum Density Fluctuations in Classical Liquids}

\author{ L.H. Ford}
 \email[Email: ]{ford@cosmos.phy.tufts.edu} 
 \affiliation{Institute of Cosmology  \\
Department of Physics and Astronomy\\ 
         Tufts University, Medford, MA 02155}
\author{N.F. Svaiter }
 \email[Email: ]{nfuxsvai@cbpf.br}
 \affiliation{Centro Brasiliero de Pesquisas Fisicas CBPF \\
 Rua Dr. Xavier Sigaud 150 \\
Rio de Janeiro, RJ, 22290 180, Brazil}

\begin{abstract}
We discuss the density fluctuations of a fluid due to zero point
motion. These can be regarded as density fluctuations in the phonon
vacuum state. We assume a linear dispersion relation with a fixed 
speed of sound and calculate the density correlation function.
We note that this function has the same form as the correlation
function for the time derivative of a relativistic massless scalar
field, but with the speed of light replaced by the speed of sound.
As a result, the study of density fluctuations in a fluid can be a
useful analog model for better understanding fluctuations in
relativistic quantum field theory. We next calculate the differential
cross section for light scattering by the zero point density 
fluctuations, and find a result proportional to the fifth power
of the light frequency. This can be understood as the product of
fourth power dependence of the usual Rayleigh cross section with
the linear frequency dependence of the spectrum of zero point
density fluctuations. We give some estimates of the relative
magnitude of this effect compared to the scattering by thermal
density fluctuations, and find that it can be of order $0.5\%$
for water at room temperature and optical frequencies. This
relative magnitude is proportional to frequency and inversely
proportional to temperature. Although the scattering by zero point
density fluctuation is small, it may be observable.
\end{abstract}


\maketitle 
 
\baselineskip=14pt

Zero point motion is a well established phenomenon, both in condensed
matter physics and relativistic quantum field theory. One effect
of the zero point motion of atoms is that the
Debye-Waller factor does not approach unity at zero temperature, but
is typically about $0.9$ at low temperature~\cite{Kittel}. This means
that the intensity of X-ray diffraction lines is reduced by
approximately $10 \%$ as a result of atomic zero point motion. In
quantum field theory, an example of a zero point phenomenon is the
Casimir effect~\cite{Casimir}, the force of attraction of bodies due 
to shifts in electromagnetic zero point energy. There is an analog
of the Casimir effect in which zero point fluctuations of the 
phonon field in a fluid also produce an analogous force~\cite{DLP}.
Unfortunately, the phononic analog is smaller than the electromagnetic
effect by the ratio of the speed of sound to the speed of light,
and is hence very small. Some authors have discussed the possibility
of an ``acoustic Casimir'' effect produced not by zero point
fluctuations, but rather by a thermal or stochastic bath of 
sound~\cite{Larraza,Bschorr}. Other authors have recently discussed
acoustic analogs of the Casimir effect in thin films~\cite{SU02}
and in Bose-Einstein condensates~\cite{RP05} or other quantum 
liquids~\cite{Recati}. It has recently been argued that the thermal
effects in liquid helium may be large enough to observe~\cite{Lamoreaux}.
 In the present paper, we will be primarily
concerned with local density fluctuations in the phonon vacuum state.
We also deal with classical as opposed to quantum liquids, so the 
the deBroglie wavelength of the atoms is small compared to the
interatomic separation.
 
Here we consider the quantization of sound waves in a fluid with a
linear dispersion relation, $\Omega_q = c_S\, q$, where  $\Omega_q$
is the phonon angular frequency, $q$ is the magnitude of the wave
vector, and $c_S$ is the
speed of sound in the fluid. This should be a good approximation for
wavelengths much longer the interatomic separation.  
Let $\rho_0$ be the mean mass density
of the fluid. Then the variation in density around this mean value is
represented by a quantum operator, $\hat{\rho}(\mathbf{x},t)$, which
may be expanded in terms of phonon annihilation and creation operators
as~\cite{LL-ST}
\begin{equation}
 \hat{\rho}(\mathbf{x},t) = \sum_{\mathbf{q}} (b_{\mathbf{q}}
 f_{\mathbf{q}} + b^\dagger_{\mathbf{q}}  f^*_{\mathbf{q}}) \,,
\end{equation}
where
\begin{equation}
 f_{\mathbf{q}} = \sqrt{\frac{\hbar \omega \rho_0}{2 V c_S^2}}
\; {\rm e}^{i(\mathbf{q} \cdot \mathbf{x} -\Omega_q\, t)} \,. 
\label{eq:mode_fnt}
\end{equation}
Here $V$ is a quantization volume. The normalization factor in 
Eq.~(\ref{eq:mode_fnt}) can be fixed by requiring that the zero point energy
of each mode be $\frac{1}{2} \hbar \Omega_q$ and using the expression for the 
energy density in a sound wave,
\begin{equation}
U = \frac{c_S^2}{\rho_0}\, {\hat{\rho}}^2\,.
\end{equation}
 In the limit in which $V
\rightarrow \infty$, we may write the density correlation function
as
\begin{equation}
\langle \hat{\rho}(\mathbf{x},t)\, \hat{\rho}(\mathbf{x}',t') \rangle
= \frac {\hbar \rho_0}{16 \pi^3 c_S^2}\, \int d^3q\, \Omega_q\,
 {\rm e}^{i(\mathbf{q} \cdot\Delta \mathbf{x} -\Omega_q\, \Delta t)}\,, 
                        \label{eq:rhorho}
\end{equation}
where $\Delta \mathbf{x} = \mathbf{x} - \mathbf{x}'$ and
$\Delta t = t-t'$. The integral may be evaluated to write the
coordinate space correlation function as
\begin{equation}
\langle \hat{\rho}(\mathbf{x},t)\, \hat{\rho}(\mathbf{x}',t') \rangle
= -\frac{\hbar \rho_0}{2 \pi^2 c_S}\; 
\frac{\Delta \mathbf{x}^2 +3 c_S^2 \Delta t^2}
{(\Delta \mathbf{x}^2 -3 c_S^2 \Delta t^2)^3}\,.
\end{equation}

This is of the same form as the correlation function for the time
derivative of a massless scalar field in relativistic quantum field
theory, 
$\langle \dot{\varphi}(\mathbf{x},t)\, \dot{\varphi}(\mathbf{x}',t')
\rangle$.
Apart from a factor of $\rho_0$, these two quantities may be obtained
from one another by interchanging the speed of light $c$ and the speed of
sound $c_S$. If $c \rightarrow c_S$, then 
\begin{equation}
\langle \dot{\varphi}(\mathbf{x},t)\, \dot{\varphi}(\mathbf{x}',t')
\rangle \rightarrow \rho_0 \,
\langle \hat{\rho}(\mathbf{x},t)\, \hat{\rho}(\mathbf{x}',t')
\rangle\,.
\end{equation}

In the limit of equal times, the density correlation function becomes
\begin{equation}
\langle \hat{\rho}(\mathbf{x},t)\, \hat{\rho}(\mathbf{x}',t) \rangle
= -\frac{\hbar \rho_0}{2 \pi^2 c_S \, (\Delta \mathbf{x})^4} \,.
\label{eq:rho=t}
\end{equation}
Thus the density fluctuations increase as $|\Delta \mathbf{x}|$
decreases. Of course, the continuum description of the fluid and the
linear dispersion relation both fail as  $|\Delta \mathbf{x}|$
approaches the interatomic separation. Also note the minus sign in
Eq.~(\ref{eq:rho=t}). This implies that density fluctuations at
different locations at equal times are anticorrelated. By contrast,
when  $c_S|\Delta t| > |\Delta \mathbf{x}|$, then
$\langle \hat{\rho}(\mathbf{x},t)\, \hat{\rho}(\mathbf{x}',t) \rangle >0$
and the fluctuations are
positively correlated. This is complete analogy with the situation in
the relativistic theory. Fluctuations inside the lightcone can
propagate causally and tend to be positively correlated. Fluctuations 
in a fluid 
for which $c_S|\Delta t| < |\Delta \mathbf{x}|$ cannot have propagated
from one point to the other, and are anti-correlated. This can be
understood physically because an over density of fluid at one point
in space requires an under density at a nearby point.

Thus the quantum density fluctuations in a fluid can serve as an analog
model for fluctuations in quantum field theory. The effect of
boundaries on the density fluctuations is similar to the effect
of reflecting boundaries on the mean squared electric and magnetic
fields. The effects of various types of boundaries on the density
fluctuations will be treated in detail in another paper~\cite{FS08},
but as an example we here quote the result for a planar boundary.  
At an impenetrable boundary, the normal derivative of the fluid density
must vanish. This is corresponds to a massless
scalar field which satisfies Neumann boundary conditions. We can
define the shift in the mean squared density, 
$\langle (\delta \rho)^2 \rangle_R$ as
\begin{equation}
\langle (\delta \rho)^2 \rangle_R = \lim_{\Delta t \rightarrow 0,
\Delta \mathbf{x}  \rightarrow 0}\; \left[
\langle \hat{\rho}(\mathbf{x},t)\, \hat{\rho}(\mathbf{x}',t')
\rangle_B
-\langle \hat{\rho}(\mathbf{x},t)\, \hat{\rho}(\mathbf{x}',t')
\rangle \right] \,,
\end{equation} 
where $\langle\; \rangle_B$ denotes the correlation function in
the presence of a boundary. At a distance $z$ from a planar boundary,
we find
\begin{equation}
\langle (\delta \rho)^2 \rangle_R = 
- \frac{\hbar \rho_0}{32 \pi^2\, c_S\, z^4}\,.
\end{equation}
Note the minus sign, which indicates a reduction in fluctuations
near the boundary.
This is analogous to the shift in the mean squared electric and
magnetic fields near a perfectly reflecting plate~\cite{SF02}
 (Electromagnetic quantities will in Lorentz-Heaviside units
throughout this paper.)
\begin{equation}
\langle E^{2}\rangle = -\langle B^{2}\rangle =
 \frac{3\hbar c}{16\pi^{2}\, z^4}\,.
\end{equation}

Next we turn to the question of whether these zero point density
fluctuations are observable. Two means for detecting density
fluctuations are by light scattering or by neutron scattering. 
We consider only the former here. The scattering of light by thermal
fluctuations has been extensively studied~\cite{HL78} in the past.
One approach utilizes the Maxwell equations with a fluctuating
dielectric function.  Write the dielectric function of the fluid
as $\epsilon = \epsilon_0 + \epsilon_1$, with $ \epsilon_0$ being 
the mean dielectric constant of the fluid, and $\epsilon_1$ the
fluctuating part. We assume that the
dielectric function is proportional to density, so we can write
\begin{equation}
\epsilon_1(\mathbf{x},t) =   \frac{\epsilon_0}{\rho_0}\,
 \hat{\rho}(\mathbf{x},t) \,.
\end{equation}
One may then use results such as  Eq.~(1.68) of Ref.~\cite{HL78}
to obtain the scattering cross section. However, here we will
summarize a different approach which leads to the same result.

Consider the quantized electromagnetic field in a non-dispersive
dielectric with dielectric constant $\epsilon_0$. The Hamiltonian
may be written in terms of the electric and magnetic fields as
\begin{equation}
H_0 = \frac{1}{2}\, \int d^3x \,( \epsilon_0 {{\mathbf E}^2
+ \mathbf B}^2 )\,.
\end{equation}
Here the electric field operator may be expanded in photon
annihilation and creation operators as
\begin{equation}
{\mathbf E}(\mathbf{x},t) = \sum_{{\mathbf k},\lambda} 
\sqrt{\frac{\hbar \omega}{2 V \epsilon_0}}\; \left[
a_{{\mathbf k},\lambda}\, \mathbf{\hat{e}}_{{\mathbf k},\lambda}\,
 {\rm e}^{i(\mathbf{k} \cdot \mathbf{x} -\omega\, t)} +
a^\dagger_{{\mathbf k},\lambda}\, \mathbf{\hat{e}}_{{\mathbf k},\lambda}
 {\rm e}^{-i(\mathbf{k} \cdot \mathbf{x} -\omega\, t)} \right]\,,
\end{equation}
where $ \mathbf{\hat{e}}_{{\mathbf k},\lambda}$ are real polarization
vectors and $\lambda$ labels linear polarization states. Here
\begin{equation}
\omega = \frac{c}{\sqrt{\epsilon_0}}\, k \,.
\end{equation}
Suppose that the electromagnetic field is coupled to the
dielectric fluctuations by the interaction Hamiltonian
\begin{equation}
H' = \frac{1}{2}\, \int d^3x \; \epsilon_1(\mathbf{x},t)\;
{\mathbf E}^2(\mathbf{x},t) \,.
\end{equation}

 We wish to calculate the amplitude for a photon in an initial
state $({\mathbf k},\lambda)$ to scatter into state  
$({\mathbf k}',\lambda')$ with the emission of a phonon into mode
$\mathbf{q}$. Thus the initial state of the photon + phonon system
is $|\psi_i \rangle = |1_{{\mathbf k},\lambda},0_\mathbf{q} \rangle$,
and the final state is 
$|\psi_f \rangle = |1_{{\mathbf k}',\lambda'},1_\mathbf{q} \rangle$.
We use first order perturbation theory and write
\begin{equation}
\langle \psi_f|H'|\psi_i \rangle =
\sqrt{\frac{\hbar^3 \omega \omega' \Omega_q}{8 V \rho_0 c_S^2}}\;
(\mathbf{\hat{e}}_{{\mathbf k},\lambda} \cdot 
\mathbf{\hat{e}}_{{\mathbf k}',\lambda'})\;
\delta_{\mathbf{k},\mathbf{k}'+\mathbf{q}}\;
{\rm e}^{i(\omega' +\Omega_q-\omega)t} \,.
\end{equation}
The transition rate is given by the usual relation
\begin{equation}
{\cal W} = \frac{2\pi}{\hbar}\, |\langle \psi_f|H'|\psi_i \rangle|\;
\rho_f\,,
\end{equation}
where the density of photon final states in energy is here given by
\begin{equation}
\rho_f = 
\frac{V\, (\omega')^2 \, \epsilon_0^\frac{3}{2}}{\hbar (2\pi c)^3}\;
d\Omega
\end{equation}
for scattering into solid angle $d\Omega$.
The incident flux of photons is given by $c/(V\, \sqrt{\epsilon_0})$.
This leads to the result for the photon scattering cross section
by zero point fluctuations
 \begin{equation}
\left(\frac{d\sigma}{d\Omega}\right)_{ZP} = 
\frac{\hbar \omega\,(\omega')^3\,\Omega_q\, {\cal V}\, \eta^4}{32
  \pi^2 c^4 c_S^2 \rho_0}\;
(\mathbf{\hat{e}}_{{\mathbf k},\lambda} \cdot 
\mathbf{\hat{e}}_{{\mathbf k}',\lambda'})^2    \,, 
\label{eq:cross_ZP}
\end{equation}
where $\cal{V}$ is the scattering volume and $\eta \sqrt{\epsilon_0}$
is the fluid's index of refraction. This relation may also be derived
from classical electromagnetic theory with a fluctuating dielectric.

The conservation of energy and momentum require that
 \begin{equation}
\omega = \omega' + \Omega_q
\end{equation}
and that
\begin{equation}
\mathbf{k} = \mathbf{k}'+\mathbf{q}
\end{equation}
The frequency of the created phonon, $ \Omega_q$, is small compared
to the light frequency $\omega$, so that $\omega' \approx \omega$
and one may show that
\begin{equation}
\Omega_q \approx \sqrt{2(1-\cos \theta)}\;\frac{c_S}{c}\; \omega\,,
\end{equation}
where $\theta$ is the scattering angle. We can now write the cross
section as
 \begin{equation}
\left(\frac{d\sigma}{d\Omega}\right)_{ZP} = \sqrt{2(1-\cos \theta)}\;
\frac{\hbar \omega^5\, {\cal V}\, \eta^4}{32
  \pi^2 c^5 c_S\rho_0}\;
(\mathbf{\hat{e}}_{{\mathbf k},\lambda} \cdot 
\mathbf{\hat{e}}_{{\mathbf k}',\lambda'})^2    \,, 
\label{eq:cross_ZP2}
\end{equation}
 The $\omega^5$ dependence
of the scattering cross section can be viewed as the product of the
$\omega^4$ dependence of Rayleigh-Brillouin scattering and one power
of $\Omega_q$, and hence of $\omega$, coming from the spectrum of
zero point fluctuations in the fluid.  Because light travels through 
the fluid at speeds much greater than the sound speed, light
scattering  reveals a nearly static distribution of density
fluctuations. Thus we can regard Eq.~(\ref{eq:cross_ZP2}) as a
probe of the fluctuations described by Eq.~(\ref{eq:rho=t}).

The scattering by  zero point fluctuations is inelastic, with
the creation of a phonon. Thus, the scattering described by
Eq.~(\ref{eq:cross_ZP2}) is really Brillouin rather than Rayleigh scattering.
However, the result is similar to that for Brillouin and Rayleigh scattering by
thermal density fluctuation is a fluid, for which the cross section 
at temperature $T$ is [See, for example Eq.~(8.3) of Ref.~\cite{HL78}.] 
 \begin{equation}
\left(\frac{d\sigma}{d\Omega}\right)_{T} = 
\frac{\omega^4\, {\cal V}\, k_B T}{16 \pi^2 c^4} \; \left[
\beta_S \rho_0^2 {\left(\frac{\partial \epsilon}{\partial
  \rho_0}\right)_S}^2 + \frac{T}{\rho_0 C_P}\,
{\left(\frac{\partial \epsilon}{\partial  T}\right)_P}^2 \right] \; 
(\mathbf{\hat{e}}_{{\mathbf k},\lambda} \cdot 
\mathbf{\hat{e}}_{{\mathbf k}',\lambda'})^2
 \,.  \label{eq:cross_T}
\end{equation}
Here $k_B$ is Boltzmann's constant, 
$\beta_S$ is adiabatic compressibility, and $C_P$ is the heat capacity
per unit mass at constant pressure.  The subscripts $S$ and $P$ refer
to derivatives of the dielectric function at constant entropy and
constant pressure, respectively. The two terms on the right hand
side of Eq.~(\ref{eq:cross_T}) have distinct physical
interpretations. The first gives the cross section for Brillouin
scattering, inelastic scattering involving either the emission or
 absorption of a phonon, leading to the Stokes and anti-Stokes lines,
respectively. The second term gives the cross section for Rayleigh
scattering. Because the  scattering by zero point density
fluctuations involves the emission of a phonon, it will contribute to
the Stokes line. Thus, we should compare Eq.~(\ref{eq:cross_ZP2}) with
the Brillouin scattering cross section
 \begin{equation}
\left(\frac{d\sigma}{d\Omega}\right)_{TB} = 
\frac{\omega^4\, {\cal V}\, k_B T}{16 \pi^2 c^4 \,c_S^2\, \rho_0} \; \left[
 \rho_0\, \left(\frac{\partial \epsilon}{\partial
  \rho_0}\right)_S \right]^2  \; 
(\mathbf{\hat{e}}_{{\mathbf k},\lambda} \cdot 
\mathbf{\hat{e}}_{{\mathbf k}',\lambda'})^2
\,,  \label{eq:cross_BT}
\end{equation}
where we have used the relation
 \begin{equation}
\beta_S= \frac{1}{\rho_0\, c_S^2}\,.
\end{equation}
The ratio of the zero point and thermal Brillouin cross sections can
be written as
 \begin{equation}
R \equiv \frac{(d\sigma/d\Omega)_{ZP}}{(d\sigma/d\Omega)_{TB}} =
 \sqrt{2(1-\cos \theta)}\,
\left(\frac{\hbar \omega}{2 k_B T}\right)\,
\left(\frac{c_S}{c}\right)\, \eta^4\, \left[
 \rho_0\, \left(\frac{\partial \epsilon}{\partial
  \rho_0}\right)_S \right]^{-2}\,.
\end{equation}   
The index of refraction, $\eta$, and the quantity 
$ \rho_0\, \left({\partial \epsilon}/{\partial  \rho_0}\right)_S$
are both of order unity, so $R$ is primarily determined by the ratio of the
photon energy to the thermal energy, and the ratio of the speed of
sound to the speed of light.

As an example, consider the case of water at room temperature and
violet light with a wavelength of $\lambda = 350nm$. In this case,
we have $c_S = 1480 m/s$ and $\eta =1.4$~\cite{CRC}. In addition, 
$ \rho_0\, \left({\partial \epsilon}/{\partial  \rho_0}\right)_S =
0.79$ ~\cite {CG66}. For back scattering, $\cos \theta = -1$,
this leads to $R \approx 0.005$. Consequently, about $0.5\%$
of the Stokes line is due to zero point motion effects. Although this
is a small fraction, it may be detectable, and will increase at lower
temperatures and shorter wavelengths. 

In summary, we have argued that the zero point density fluctuations
in a fluid are of interest both as an analog model for fluctuations
in relativistic quantum field theory, and in their own right. These
fluctuations are potentially observable in light scattering
experiments.

\vspace{1cm}

\begin{acknowledgments}
This work was supported in part by the National Science Foundation
under Grant PHY-0555754 and by Conselho Nacional de Desenvolvimento
Cientifico e Tecnologico do Brasil (CNPq).
\end{acknowledgments}

\end{document}